\documentclass[12pt]{article}

\newcommand{\bea}{\begin{eqnarray}}
\newcommand{\eea}{\end{eqnarray}}

\title{
\textbf{Three-point Green function of massless QED in position space to lowest order}}
\author{
Indrajit Mitra\footnote{E-mail: imitra@iitk.ac.in}\\\\
Department of Physics, Indian Institute of Technology,\\ 
Kanpur 208016, India}
\date{}
\begin{document}
\maketitle
\begin{abstract}
The transverse part of the three-point Green function
of massless QED is determined to the lowest order in position space.
Taken together with the evaluation of the longitudinal part in
arXiv:0803.2630, this gives a relation for QED which is
analogous to the star-triangle relation. We relate our result to
conformal-invariant three-point functions.
\end{abstract}

\section{Introduction}
\label{intro}
In theories of massless particles with dimensionless couplings, tree-level
integrals in position space can often be evaluated exactly. The simplest example,
which evaluates the three-point Green function involving three massless scalar fields
\cite{deramo, sym, im} is called the star-triangle relation (also called
the uniqueness relation). A similar relation
has been found for the massless Yukawa theory as well \cite{sym, im}. For massless
QED with dimensionless coupling in general number of dimensions, the longitudinal part
of the three-point function
to the lowest order in position space was recently determined in Ref.\ \cite{im}.
In this paper, we determine the transverse part. Taken together, these two results
give a relation analogous to the star-triangle relation.

The star-triangle relation, being exact, is very useful in higher order calculations
in perturbation theory \cite{kaza, vas, cvetic}.
The considerations of the
present work are directly relevant for
massless QED$_3$, as this theory 
has a dimensionless effective coupling constant in the infrared \cite{appel, mrs}.
So, application to higher orders of this theory
is the main motivation for our work.

Moreover, as explained in the Introduction of Ref.\ \cite{im}, our result is expected to
exhibit conformal-invariant structures. The investigation of conformal transformation of the
gauge field in Refs.\ \cite{palchik, comm} led to the conformal-invariant structures
of the vertex function of QED. It is therefore of interest to find out whether an explicit
lowest order calculation brings out these structures.

The paper is organized as follows. In Sec.\ \ref{trans}, the transverse part of the 
three-point function of QED is evaluated.
In Sec.\ \ref{tot}, we give the total result, including the longitudinal part. 
In Sec.\  \ref{conf}, we relate our result to conformal-invariant three-point functions.
Our conclusions are presented in Sec.\ \ref{concl}.
In the Appendix, we perform a check of
our result for the case of $D=4$.
\section{Transverse part of the three-point Green function}
\label{trans}
As in Ref.\ \cite{im}, we will use the operator algebraic method due to Isaev \cite{isaev1}
in which
one reduces Feynman integrals to products
of position and momentum operators $\hat{q}_i$ and $\hat{p}_i$ taken between
position eigenstates.  
As explained in Sec.\ 2 of Ref.\ \cite{im}, this method involves starting from the
$``\hat{p}\hat{q}\hat{p}"$ form and passing to the
$``\hat{q}\hat{p}\hat{q}"$ form. In what follows, we proceed in the same way as in
Sec.\ 4 of Ref.\ \cite{im}, where we evaluated the longitudinal part of the 
three-point function with fermion propagator 
$\rlap/p/p^2$ and photon scale dimension
$\Delta=1$ \cite{palchik, comm} (this corresponds to both QED$_4$ and the infrared limit
of massless QED$_3$). 
We will use  
the regularized scale dimensions given by Eq.\ (16) of Ref.\ \cite{im}:
\bea
\Delta=1+\epsilon\,,~d=\frac{D-1-\epsilon}{2}\,.                      \label{reg}
\eea
Here $d$ denotes the fermion scale dimension and $D$ the number of (Euclidean) dimensions.
The starting $``\hat{p}\hat{q}\hat{p}"$ form for the transverse part of 
$\langle \psi(x)\bar\psi(0)A_k(y)\rangle$ is then
\bea
\Gamma_k^t=\gamma_i\gamma_l\gamma_j\hat{p}_i\hat{p}^{\,-2-\epsilon}
          \hat{q}_j\hat{q}^{\,-D+\epsilon}
          \hat{p}^{\,-D+2+2\epsilon}\,{\cal P}_{lk}\,,                          \label{eq:1}
\eea
where ${\cal P}_{lk}$ is the transverse projection operator:
\bea
{\cal P}_{lk}=\delta_{lk}-\hat{p}_l\hat{p}_k\hat{p}^{\,-2}\,.
\eea
Eq.\ (\ref{eq:1}) is the counterpart of Eq.\ (17) of Ref.\ \cite{im}, which gives the
longitudinal part. Thus, for $\epsilon= 0$, Eq.\ (\ref{eq:1}) gives
\bea
\langle x|\Gamma_k^t|y\rangle=i\frac{(D-2)}{(2\pi)^D}
\int d^Dz~\frac{\rlap/x-\rlap/z}{|x-z|^{D}}\,\gamma_l\,
            \frac{\rlap/z}{|z|^{D}}\,
            \Bigg(\delta_{lk}-\frac{\partial{^y_l}\partial{^y_k}}{(\partial^2)^y}\Bigg)\,
            \frac{1}{|y-z|^{2}}\,.                                               \label{eq:1a}                           
\eea
Eq.\ (\ref{eq:1a}) is the counterpart of Eq.\ (15) of Ref.\ \cite{im}.

Let us now recall from Ref.\ \cite{im} that in proceeding from Eq.\ (\ref{eq:1}) to the
$``\hat{q}\hat{p}\hat{q}"$ form, one has to move $\hat{q}_i$ (or $\hat{p}_i$)
through powers of $\hat{p}^{\,2}$ (or $\hat{q}^{\,2}$) by using
$[\hat{q}_i,\hat{p}^{\,2\alpha}]=i2\alpha\hat{p}^{\,2\alpha-2}\hat{p}_i$
(or $[\hat{p}_i,\hat{q}^{\,2\alpha}]=-i2\alpha\hat{q}^{\,2\alpha-2}\hat{q}_i$). One also has to
make use of the key relation
\bea
\hat{p}^{\,-2\alpha}\hat{q}^{\,-2(\alpha+\beta)}\hat{p}^{\,-2\beta}
=\hat{q}^{\,-2\beta}\hat{p}^{\,-2(\alpha+\beta)}\hat{q}^{\,-2\alpha}\,,              \label{key1}
\eea
which is the star-triangle relation in the operator form, 
at some intermediate stage. Using $\{\gamma_j,\gamma_l\}=2\delta_{jl}$ in Eq.\ (\ref{eq:1}), 
one can write $\Gamma_k^t$ as the sum of two terms, say, $\Gamma_k^{t(1)}$ and 
 $\Gamma_k^{t(2)}$. One term is
\bea
\Gamma_k^{t(1)}=2\gamma_i\hat{p}_i\hat{p}^{\,-2-\epsilon}
          \hat{q}^{\,-D+\epsilon}\hat{q}_l
          \hat{p}^{\,-D+2+2\epsilon}\,{\cal P}_{lk}\,.
\eea
Now $\hat{q}_l$ can be taken through $\hat{p}^{\,-D+2+2\epsilon}$ to the right without generating an extra term,
since the commutator, being proportional to $\hat{p}_l$, is annihilated by ${\cal P}_{lk}$. Next, use
Eq.\ (\ref{key1}) to obtain
\bea
\Gamma_k^{t(1)}=2\gamma_i\hat{p}_i\hat{q}^{\,-D+2+2\epsilon}
          \hat{p}^{\,-D+\epsilon}\hat{q}^{\,-2-\epsilon}
          \hat{q}_l\, {\cal P}_{lk}\,.
\eea
Finally, one takes $\hat{p}_i$ through $\hat{q}^{\,-2-\epsilon}$ to the left to arrive at
\bea
\Gamma_k^{t(1)}=2\gamma_i(\hat{q}^{\,-D+2+2\epsilon}\hat{p}_i+i(D-2-2\epsilon)
                \hat{q}^{\,-D+2\epsilon}\hat{q}_i)
                \hat{p}^{\,-D+\epsilon}\hat{q}^{\,-2-\epsilon}
                \hat{q}_l\, {\cal P}_{lk}\,.                             \label{eq:2}
\eea
The other term is
\bea
\Gamma_k^{t(2)}=-\gamma_i\gamma_j\gamma_l\hat{p}_i\hat{p}^{\,-2-\epsilon}
          \hat{q}_j\hat{q}^{\,-D+\epsilon}
          \hat{p}^{\,-D+2+2\epsilon}\,{\cal P}_{lk}\,.     \label{eq:3}           
\eea
Except for $\gamma_l$ and ${\cal P}_{lk}$, this is the 
$``\hat{p}\hat{q}\hat{p}"$ form for the vertex of the massless Yukawa 
theory (see Eq.\ (5) of Ref.\ \cite{im}). Therefore, following exactly the same
steps as for that case, Eq.\ (\ref{eq:3}) leads to
\bea
\Gamma_k^{t(2)}&=&-\gamma_i\gamma_j\gamma_l \hat{p}_i\,(\hat{q}_j\hat{p}^{\,-2-\epsilon}
                 +i(2+\epsilon)\hat{p}^{\,-4-\epsilon}\hat{p}_j)\,
                 \hat{q}^{\,-D+\epsilon}\hat{p}^{\,-D+2+2\epsilon}\,{\cal P}_{lk} 
                                                                         \label{eq:4}\\
              &=&-(\gamma_i\gamma_j\gamma_l\hat{p}_i\hat{q}_j
               \hat{q}^{\,-D+2+2\epsilon}
              \hat{p}^{\,-D+\epsilon}\hat{q}^{\,-2-\epsilon}
             +i(2+\epsilon)\gamma_l
               \hat{q}^{\,-D+2+2\epsilon}
              \hat{p}^{\,-D+\epsilon}\hat{q}^{\,-2-\epsilon})\,{\cal P}_{lk}   \\
               &=& (-\gamma_i\gamma_j\gamma_l \hat{q}_j \hat{q}^{\,-D+2+2\epsilon} 
                  \hat{p}_i \hat{p}^{\,-D+\epsilon}\hat{q}^{\,-2-\epsilon}
                 +i\epsilon \gamma_l\hat{q}^{\,-D+2+2\epsilon}
                  \hat{p}^{\,-D+\epsilon}\hat{q}^{\,-2-\epsilon})\,{\cal P}_{lk}\,.
                                                                               \label{eq:5}
\eea
It may be noted that though the matrix-element of $\hat{p}^{\,-4-\epsilon}$ in position space
is infrared-divergent for $D\le 4$, in
Eq.\ (\ref{eq:4}) this operator comes multiplied with $\gamma_i\gamma_j\hat{p}_i\hat{p}_j$ 
and gives $\hat{p}^{\,-2-\epsilon}$. Therefore it is legitimate to use Eq.\ (\ref{key1})
in the next step (the star-triangle relation, as given by Eq.\ (1) of Ref.\ \cite{im}, 
holds only when $\delta_i > 0$: see Ref.\ \cite{sym}). 

Now,
\bea
\langle x|\Gamma_k^t|y\rangle=\Bigg(\delta_{kl}-
                             \frac{\partial{^y_k}\partial{^y_l}}{(\partial^2)^y}\Bigg)
                             \langle x|(\Gamma{_l^{t(1)}}'+\Gamma{_l^{t(2)}}')|y\rangle\,,
                                                             \label{eq:6}
\eea
where $\Gamma{_l^{t(1)}}'$ and $\Gamma{_l^{t(2)}}'$ are given by the
right-hand sides of Eqs.\ (\ref{eq:2})
and (\ref{eq:5}), {\it omitting} ${\cal P}_{lk}$. It is important to note that
due to the presence of the transverse projection operator
on the  right-hand side of Eq.\ (\ref{eq:6}),
{\it if any term in the matrix element following 
it can be expressed as a derivative with respect to $y_l$,
that term will not contribute.}
The evaluation of the matrix element is to be done using the matrix elements
given in the Appendix of Ref.\ \cite{im}.
It is found that in three of the resulting terms, the diverging
$\Gamma(\epsilon/2)$ comes multiplied by $\epsilon$. So
taking $\epsilon\rightarrow 0$
gives finite results for these terms in a straightforward way.
However, there is also the contribution (this is the second term on 
the right-hand side of Eq.\ (\ref{eq:2}),
without ${\cal P}_{lk}$, taken between $\langle x|$ and $|y\rangle$):
\bea
2i\frac{(D-2-2\epsilon)\Gamma(\epsilon/2)}{\pi^{D/2}2^{D-\epsilon}\Gamma(D/2-\epsilon/2)}\,
                 \frac{\rlap/x}{|x|^{D-2\epsilon}}\,
                 \frac{1}{|x-y|^\epsilon}\,
                 \frac{y_l}{|y|^{2+\epsilon}}\,.                                              \label{eq:6a}
\eea
For  $\epsilon\rightarrow 0$, there is an apparent $O(1/\epsilon)$ singularity.
But the  $y$-dependence of this singular term is just $y_l/|y|^2$, which equals $(\partial/\partial y_l)\ln|y|$,
and so the singular term will not contribute. 
There is still a finite contribution from (\ref{eq:6a}), which has to
be taken into account. The $y$-dependences in this contribution  will involve
only $y_l/|y|^2$,  and also
$y_l/|y|^2$ times 
\mbox{$\ln(|x-y||y|/|x|^2)$}. The only surviving contribution is
$(y_l/|y|^2)\ln|x-y|$. 
(Since $(y_l/|y|^2)\ln|y|=(1/2)(\partial/\partial y_l)(\ln|y|)^2$,
it drops out.) Putting everything together, we arrive at
\bea
\langle x|(\Gamma{_l^{t(1)}}'\!\!+\Gamma{_l^{t(2)}}')|y\rangle=
\frac{i}{\pi^{D/2}2^{D-1}\Gamma(D/2)}\,
\Bigg(\frac{(\rlap/x-\rlap/y)\gamma_l\rlap/y}{|x|^{D-2} |x-y|^2 |y|^2}
      -2(D-2)\frac{\ln|x-y|\rlap/x y_l}{|x|^D |y|^2}\Bigg)\,. 
                                          \label{eq:7}
\eea
The right-hand sides of Eqs.\ (\ref{eq:1a}), (\ref{eq:6}) and (\ref{eq:7}) taken together
give the result for the transverse part of the three-point function. 
We reiterate that we have the logarithm of the dimensionful object $|x-y|$ in Eq.\ (\ref{eq:7})
only because we have dropped all terms annihilated by the transverse projection operator. 
\section{Total result for three-point Green function}
\label{tot}
Adding the longitudinal part given by Eq.\ (28) of Ref.\ \cite{im}, the result for
$\langle T(\psi(x)\bar\psi(0)A_k(y)\rangle$ to the lowest order is
\bea
&&\int d^Dz~\frac{\rlap/x-\rlap/z}{|x-z|^{D}}\,\gamma_l\,
            \frac{\rlap/z}{|z|^{D}}\,
            \Bigg(\delta_{kl}-(1-\eta)\frac{\partial{^y_k}\partial{^y_l}}{(\partial^2)^y}\Bigg)\,
            \frac{1}{|y-z|^{2}}                                           \nonumber\\
&=&\frac{2\pi^{D/2}}{(D-2)\Gamma(D/2)}\,\Bigg[
   \Bigg(\delta_{kl}-\frac{\partial{^y_k}\partial{^y_l}}{(\partial^2)^y}\Bigg)
   \Bigg(\frac{(\rlap/x-\rlap/y)\gamma_l\rlap/y}{|x|^{D-2} |x-y|^2 |y|^2}
      -2(D-2)\frac{\ln|x-y|\rlap/x y_l}{|x|^D |y|^2}\Bigg)                   \nonumber\\
&&      +\eta\frac{\rlap/x}{|x|^D}
                \Bigg(\frac{(x-y)_k}{|x-y|^2}+\frac{y_k}{|y|^2}\Bigg)\Bigg]                \label{total}
\eea
Here $\eta$ is the gauge parameter. Letting
$x=x_1-x_2$ and  $y=x_3-x_2$, and also changing to a new integration
variable $x_4$ defined by $z=x_4-x_2$, one can immediately write this relation in a
manifestly translation-invariant form, as was done for the relations in Ref.\ \cite{im}.
The right-hand side then depends on the differences of the three external coordinates, taken in pairs.
Thus we have a relation for massless QED similar to the star-triangle relation,
with the not only the structure functions but also their coefficients exactly determined. 
That the transverse projection
operator is present on the right-hand side of Eq.\ (\ref{total}) 
is not a cause for concern, because
its presence is actually useful for relating to the conformal-invariant structures given in the literature, and also
for calculation at higher orders (as the orthogonality of the transverse part and the
remaining, longitudinal part is manifest).

In Sec.\ \ref{conf}, we compare Eq.\ (\ref{total}) with the structure functions given
from general considerations of conformal invariance in
Refs.\ \cite{palchik} and \cite{comm}. 
We find that the logarithmic term in the transverse part does not appear in these references.
Because of this, we perform a check in the  Appendix to confirm the presence of this term.
It is based on the observation that $(\partial^2)^y$ acting on the right-hand side of 
Eq.\ (\ref{total}) can be readily calculated. Also, for $\eta=1$, the $y$-dependence
of the left-hand side is just $1/|y-z|^{2}$ and for $D=4$, $(\partial^2)^y$ acting on this
gives a delta function; so $(\partial^2)^y$ acting on 
the left-hand side can be evaluated as well. The two resulting expressions are then found to
match only if the logarithmic term is present. 
\section{Relation to conformal-invariant functions}
\label{conf}
Of the transformations associated with the conformal algebra, 
invariance (or covariance) of our result under translation, (Euclidean) rotation and scaling 
is obvious. Our concern will therefore be with the effect of the coordinate inversion $R$:
$(Rx)_\mu\equiv x_\mu/x^2$. We first summarize some results on $R$ which will
be directly useful for our purpose.

A conformal vector with scale dimension $\Delta$ transforms under $R$ as \cite{ferrara}
\bea
A_k(x)\rightarrow UA_k(x)=|x|^{-2\Delta}(\delta_{kl}-2x_kx_l/x^2)A_l(Rx)\,.         \label{law1}
\eea
With this transformation law, the two conformal-invariant structures for 
$\langle \psi(x)\bar\psi(0)A_k(y)\rangle$ are
($d$ being the scale dimension of the fermion)
\bea
C^{d,\Delta}_{1k}(x,y)&=&
             \frac{(\rlap/x-\rlap/y)\gamma_k\rlap/y}{|x|^{2d-\Delta} |x-y|^{\Delta+1} |y|^{\Delta+1}}\,,
                                                                             \label{eq:18}\\
C^{d,\Delta}_{2k}(x,y)&=&
             \frac{\rlap/x}{|x|^{2d-\Delta+2} |x-y|^{\Delta-1} |y|^{\Delta-1}}
               \Bigg(\frac{(x-y)_k}{|x-y|^2}+\frac{y_k}{|y|^2}\Bigg)             \label{eq:18a}\\
       &=&\frac{\rlap/x}{|x|^{2d-\Delta+2} |x-y|^{\Delta-1} |y|^{\Delta-1}}
           \partial{^y_k}\ln\frac{|y|}{|x-y|}\,.                                \label{eq:19}
\eea
(See Ref.\ \cite{book}; the structures for  $\Delta=1$
are given in Refs.\ \cite{palchik} and \cite{comm}.) 

For $\Delta=1$, Eq.\ (\ref{law1}) leads to a conformal-invariant propagator 
without a transverse part. 
In order to accommodate the usual covariant-gauge propagator
(as used by us), Eq.\ (\ref{law1}) can be modified
to the following {\it new} transformation law \cite{palchik, comm}:
\bea
A_k(x)\rightarrow {\tilde U}A_k(x)=[(1-P)U(1-P)+UP]A_k(x)                        \label{law2}
\eea
Here $P=\partial_k\partial_l/\partial^2$ is the longitudinal projection operator.
The new conformal-invariant structures for
$\langle \psi(x)\bar\psi(0)A_k(y)\rangle$ are
then \cite{palchik, comm} $C^{d,\Delta=1}_{2k}(x,y)$ , which is purely longitudinal in $y$
(see Eq.\ (\ref{eq:19})), and
$(\delta_{kl}-\partial{^y_k}\partial{^y_l}/(\partial^2)^y)C^{d,\Delta=1}_{1l}(x,y)$,
which is (manifestly) transverse.
On putting $d=(D-1)/2$, we see that {\it these are precisely the terms present in our result of
Eq.\ (\ref{total}), except for the logarithmic term.}

To understand the logarithmic term, we first note that this term, which came from
$(\delta_{kl}-\partial{^y_k}\partial{^y_l}/(\partial^2)^y)$ operating on
the expression (\ref{eq:6a}), also emerges from
$(\delta_{kl}-\partial{^y_k}\partial{^y_l}/(\partial^2)^y)$ operating on
\bea
2i\frac{(D-2-2\epsilon)\Gamma(\epsilon/2)}{\pi^{D/2}2^{D-\epsilon}\Gamma(D/2-\epsilon/2)}\,
                 \frac{\rlap/x}{|x|^{D-2\epsilon}}\,
                 \frac{1}{|y|^\epsilon}\,
                 \frac{(x-y)_l}{|x-y|^{2+\epsilon}}\,.                                              \label{eq:6b}
\eea
The reason is that, after taking $\epsilon\rightarrow 0$ and discarding all terms which are
derivatives with respect to $y_l$, the surviving term has the $y$-dependence
$\ln|y|(x-y)_l/|x-y|^2$, and this is the same as $\ln|x-y|y_l/|y|^2$ (upto
$(\partial/\partial y_l)(\ln|y|\ln|x-y|)$). Therefore, the average of the expressions
(\ref{eq:6a}) and (\ref{eq:6b}) also gives our logarithmic term. Comparing with Eq.\ (\ref{eq:18a}),
we find that this average is
$C^{d,\Delta}_{2l}(x,y)$ {\it with $d$ and $\Delta$ given by the regularized
scale dimensions of Eq.\ (\ref{reg})}.

So the logarithmic term arises in the following way. The function $C^{d,\Delta}_{2k}(x,y)$ is longitudinal
for  $\Delta=1$, but for $\Delta=1+\epsilon$, it develops an $O(\epsilon)$ part which is not
longitudinal and therefore not annihilated by the transverse projection operator. Because 
of the presence of
a coefficient of $O(1/\epsilon)$, this leads to a contribution which survives the limit
$\epsilon\rightarrow 0$.

We now briefly address the question: is the
logarithmic term invariant under the new transformation law given in Eq.\ (\ref{law2})?
To answer this, let us first go through the demonstration of invariance of a transverse
three-point function , say $\Gamma_k\equiv (1-P)C_k$, under Eq.\ (\ref{law2}) for $\Delta=1$.
This function transforms to $\Gamma'_k=(1-P)U^{\Delta=1}(1-P)C_k$, since $P(1-P)=0$
and $(1-P)^2=1-P$. (We do not write the transformation of the spinors, which can be
taken into account trivially). Next, using $PU^{\Delta=1}P=U^{\Delta=1}P$ (see Ref.\ \cite{palchik}), 
we get $\Gamma'_k=(1-P)U^{\Delta=1} C_k$.
Finally, if $C_k$ satisfies $U^{\Delta=1} C_k=C_k$, we arrive at $\Gamma'_k=\Gamma_k$. But 
$C_{2k}^{\Delta=1+\epsilon}$ satisfies
$U^{\Delta=1+\epsilon} C_k=C_k$,
and since 
$PU^{\Delta=1+\epsilon}P=U^{\Delta=1+\epsilon}P$ does not hold, 
the demonstration
outlined above cannot be carried out
for $\Gamma_k=(1-P)C_{2k}^{\Delta=1+\epsilon}$.
A direct calculation with the logarithmic term also
confirms non-invariance under Eq.\ (\ref{law2}).

This non-invariance is, however, not an artifact of the regularization of the scale dimensions,
as the check performed in the Appendix does not use the regularization at any stage.
The possible reason for this non-invariance is that invariance under the
the new transformation given by Eq.\ (\ref{law2}) can be realized only non-perturbatively
and not in perturbation theory, as stated in Ref.\ \cite{eprint}.
\section{Conclusion}
\label{concl}
In this work, we have completed the evaluation of 
the three-point Green function
of massless QED to the lowest order in position space.
This has resulted in a relation which is analogous to the star-triangle relation,
and can be used for calculations to higher orders in perturbation theory.
The transverse part of the three-point function
was found to contain a logarithmic term, and the presence of this term was checked
by a calculation in $D=4$. The relation of the various terms in our result to
conformal-invariant three-point functions was explained.
\appendix
\section*{Appendix}
We consider the case of $D=4$ and Feynman gauge. Let us take Eq.\ (\ref{total}) with $\eta=1$ 
and operate on both sides with 
$(\partial^2)^y$ in $D=4$. For the left-hand side, we have
\bea
(\partial^2)^y\int d^4z~\frac{\rlap/x-\rlap/z}{|x-z|^4}\,\gamma_k\,
            \frac{\rlap/z}{|z|^4}\,
            \frac{1}{|y-z|^{2}}
            =-4\pi^2 ~\frac{\rlap/x-\rlap/y}{|x-y|^4}\,\gamma_k\,
               \frac{\rlap/y}{|y|^4}\,.                                   \label{eqleft}
\eea
For the right-hand side, the following equations are to be used:
\bea
\Big((\partial^2)^y\delta_{kl}-\partial{^y_k}\partial{^y_l}\Big)
\frac{(\rlap/x-\rlap/y)\gamma_l\rlap/y}{ |x-y|^2 |y|^2}
&=&-\frac{4(x^2(\rlap/x-\rlap/y)\gamma_k\rlap/y-y^2\rlap/x x_k-x^2\rlap/x y_k+2x\cdot y\rlap/x y_k)}
       {|x-y|^4 |y|^4}                                             \label{eqrt1} \\
\Big((\partial^2)^y\delta_{kl}-\partial{^y_k}\partial{^y_l}\Big)
\frac{\ln|x-y|y_l}{ |y|^2}
&=&\frac{x_k}{|x-y|^2|y|^2}  
 +\frac{2(x\cdot y-y^2)}{|x-y|^2|y|^2}
   \Bigg(\frac{(x-y)_k}{|x-y|^2}+\frac{y_k}{|y|^2}\Bigg)            \label{eqrt2}  \\
(\partial^2)^y\Bigg(\frac{(x-y)_k}{|x-y|^2}+\frac{y_k}{|y|^2}\Bigg)
&=&-4\Bigg(\frac{(x-y)_k}{|x-y|^4}+\frac{y_k}{|y|^4}\Bigg)\,.           \label{eqrt3}
\eea
We can now evaluate the action of $(\partial^2)^y$ operating on the right-hand side
of Eq.\ (\ref{total}). The result (with $\eta=1$) is found to match Eq.\ (\ref{eqleft}).

However, if we did not have the logarithmic term on the right-hand side
of Eq.\ (\ref{total}), the action of $(\partial^2)^y$ on it would have produced
\bea
-\frac{4\pi^2}{|x|^4 |x-y|^4 |y|^4}\!\!\!\!
  &[\!\!\!\!&\{|x|^4(\rlap/x-\rlap/y)\gamma_k\rlap/y-x^2y^2\rlap/x x_k-|x|^4\rlap/x y_k+2x\cdot y x^2 \rlap/x y_k\}
   \nonumber\\
    &&+\{|x-y|^4\rlap/x y_k +|y|^4 \rlap/x(x-y)_k\}]                \label{ex27}
\eea
Here the two expressions within the curly brackets come from
 Eqs.\ (\ref{eqrt1}) and (\ref{eqrt3}); thus they
are the contributions from the two
structure functions given in Ref.\ \cite{palchik}.
Let us now note that 
there is an  $(x\cdot y)^2 \rlap/x y_k$ term when we expand out the second expression 
within the curly brackets, and this term would remain
uncancelled in (\ref{ex27}) {\it even if we had two different coefficients
with the two expressions}.
But there is no such term from Eq.\ (\ref{eqleft}). Therefore it is impossible to express
the lowest-order  QED three-point function in terms of only the two standard structure functions
invariant under Eq.\ (\ref{law2}), and the logarithmic term is essential.

\end{document}